\documentclass[a4paper,fleqn]{cas-sc}

 \usepackage[numbers]{natbib}
\usepackage{enumitem}  
\usepackage{float}
\usepackage{graphicx}
\usepackage{tabularx}
\usepackage{ltablex}     
\usepackage{multirow}
\usepackage{booktabs}    
\usepackage{fontawesome5} 
\usepackage{pdflscape}
\usepackage{listings}
\usepackage{xcolor}
\usepackage{graphicx}
\usepackage{subcaption}
\usepackage{calc}
\usepackage{changepage}

\setlist[enumerate]{noitemsep, topsep=0pt}
\setlist[itemize]{noitemsep, topsep=0pt}

\def\tsc#1{\csdef{#1}{\textsc{\lowercase{#1}}\xspace}}
\tsc{WGM}
\tsc{QE}
\tsc{EP}
\tsc{PMS}
\tsc{BEC}
\tsc{DE}

\begin{document}
\let\WriteBookmarks\relax
\def\floatpagepagefraction{1}
\def\textpagefraction{.001}

\shorttitle{Role of Graphics in Disaster Communication}

\shortauthors{Madugalla et~al.}

\title [mode = title]{Role of Graphics in Disaster Communication: Practitioner Perspectives on Use, Challenges, and Inclusivity}                      



\author[1]{Anuradha Madugalla}

\cormark[1]

\fnmark[1]

\ead{anuradha.madugalla@deakin.edu.au}



\affiliation[1]{organization={ School of Information Technology, Deakin University},
   city={Melbourne},
     state={Victoria},
   country={Australia}}

\author[2]{Yuqing Xiao}
\author[2]{John Grundy}

\affiliation[2]{organization={Faculty of Information Technology, Monash University},
   city={Melbourne},
     state={Victoria},
   country={Australia}}

\begin{abstract}
Information graphics, such as hazard maps, evacuation diagrams, and pictorial action guides, are widely used in disaster risk communication. These visuals are important because they convey hazard information quickly, reduce reliance on lengthy text, and support decision-making in time-critical situations. However, despite their importance, disaster information graphics do not work equally well for all audiences. In practice, many graphics remain difficult to interpret, and their accessibility for vulnerable populations is still uneven and underexplored. Despite their central role, there has been little detailed empirical work examining how graphics shape disaster communication outcomes, what challenges practitioners face in using them, and, most importantly, how inclusive current disaster graphics are in real-world settings. To address this gap, we examine how information graphics are currently produced and used in disaster communication, what issues emerge in practice, and how inclusivity is understood and addressed. We conducted semi-structured interviews with disaster communication practitioners and researchers (P1–P5) to examine the role of graphics across preparedness, warning, and response contexts, as well as the barriers experienced by vulnerable communities. Our findings show that graphics are widely expected and heavily relied upon, yet significant accessibility gaps persist for groups such as people with vision impairments, older adults, and culturally and linguistically diverse communities. Participants also highlighted that inclusive adaptations are difficult to achieve during unfolding emergencies due to operational constraints, limited guidance, and resource barriers. Based on these findings, we outline recommendations for disaster management agencies and graphic designers and identify research directions for technological and adaptive support to make disaster graphics more inclusive at scale.

\end{abstract}

\begin{keywords}
Disaster risk communication \sep Information graphics \sep Inclusivity \sep Practitioner perspectives  \sep Disaster visualisation
\end{keywords}

\maketitle

\section{Introduction}
The frequency and intensity of natural hazards are increasing globally, intensifying the need for effective disaster risk communication that enables communities to understand risks and take appropriate protective actions. International frameworks such as the Sendai Framework for Disaster Risk Reduction emphasise timely, clear, and accessible communication as a cornerstone of disaster preparedness, response, and recovery \cite{SendaiFr7_online}. Within this context, visual communication has become a dominant mechanism for conveying disaster-related information, particularly in time-critical warning settings where communities must rapidly interpret what is happening and decide what to do next.

It is widely recognised across domains such as health, education, and safety communication that well-designed visuals can improve attention, comprehension, and recall, especially when audiences must process information under stress or uncertainty \cite{houts_role_2006}. This is directly applicable to disaster risk contexts. Across early warning systems, preparedness campaigns, and emergency response platforms, information graphics such as hazard maps, evacuation diagrams, icon-based alerts, and infographics are widely used to communicate risk. These visual tools are often valued because they can summarise complex information quickly, reduce reliance on lengthy textual explanations, and support communication across diverse channels, including mobile apps, websites, broadcast media, and printed materials. As disasters increasingly unfold in digitally mediated environments, information graphics have become a central part of public-facing disaster communication infrastructure.

\subsection{Motivation}
Despite their widespread use, disaster information graphics are often assumed to be universally understood. In practice, however, evidence shows that communities can misinterpret maps, warning indicators, and symbols, particularly when visuals are complex or unfamiliar. Prior work demonstrates that the design of warning messages and maps can significantly shape public response, sometimes producing confusion rather than clarity \cite{liu_is_2017}. Disaster graphics may rely on inconsistent iconography, dense information layers, or visual conventions that require specific forms of literacy. At the same time, many widely distributed preparedness materials provide simplified guidance that may not account for the diverse needs of communities, including people with disabilities, culturally and linguistically diverse groups, older adults, and those with limited digital access. As a result, disaster information graphics can unintentionally create new barriers, even when their goal is to improve understanding and safety.

\begin{figure}
  \includegraphics[width=\textwidth, trim={0cm 12cm 16cm 0cm},clip]{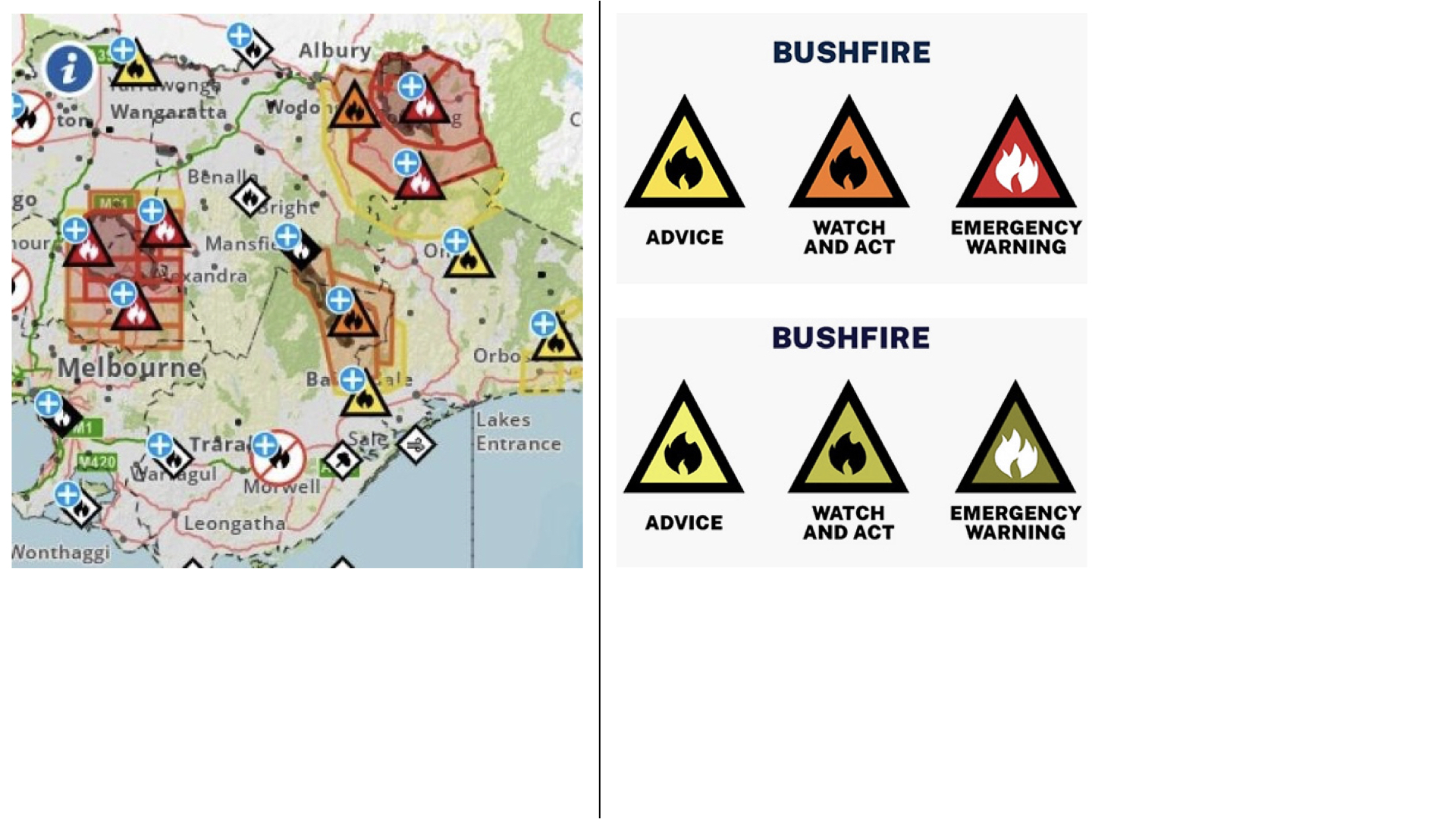}
  \caption{(a) A Complex Hazard Map, (b) Top: Bushfire Warning Symbol Colors, Bottom: Bushfire Warning Symbol Colours for Protanopia -Red Color Blindness}
  \label{motivation}
\end{figure}

Consider, for example, Figure~\ref{motivation}. On the left, a hazard map presents multiple overlapping visual layers, including hazard zones for bushfires and strong winds, as well as boundary outlines and symbolic markers. While such representations are intended to support rapid situational awareness, in this example, it may not be immediately clear which areas are most threatened, what actions are required, or how to interpret competing overlays and symbols. On the right of Figure~\ref{motivation}, colour-coded icon systems are used to convey bushfire warning levels. This colour-coding system was designed to communicate escalating urgency (e.g., \textit{Advice}, \textit{Watch and Act}, \textit{Emergency Warning}), yet its effectiveness depends on users’ ability to reliably distinguish among colour differences. For vulnerable users, such as people with Protanopia (red colour blindness), these distinctions may be difficult to access, particularly when graphics are reproduced in low-contrast formats or viewed under stress. With over 350 million people worldwide affected by colour vision deficiency, this is not a marginal concern \cite{FactsAbo79_online}. Together, these examples show the need for a deeper understanding of how disaster information graphics are currently used in practice, the challenges they introduce, and their inclusivity for diverse and vulnerable populations.

\subsection{Research Questions}
Although disaster information graphics are widely relied upon, there is still a limited understanding of their role in real-world disaster communication practice. First, while visual risk communication is frequently discussed in principle, fewer studies examine how practitioners and agencies perceive and use graphics across preparedness, warning, and response phases. Second, there is insufficient analysis of the practical barriers that shape graphic effectiveness, such as information overload, misleading cues, or mismatches between intended and received meaning. Third, while research in health communication and disaster education highlights the value of visuals, there remains little evidence explaining how specific design elements, such as colour, complexity, symbol choices, or layering, shape comprehension and action in disaster contexts \cite{houts_role_2006,yavar_effective_2012}. Finally, and most critically, there is limited empirical work examining how inclusive current disaster graphics are for vulnerable populations, despite increasing recognition of digital vulnerability and unequal access to warning information~\cite{hao_examining_2022}.

To address these gaps, this study investigates the following research questions:

RQ1: What is the role of information graphics in disaster communication?

RQ2: What are the issues in the current use of information graphics in disaster communication?

RQ3: How inclusive are these graphics for vulnerable populations?

\subsection{Study Contributions}
To answer these research questions, we conducted semi-structured interviews with disaster communication practitioners and researchers (P1--P5). Based on findings from these discussions, this study makes the following contributions.

\noindent\textit{C1: Empirical insight into real-world disaster graphics: }We provide one of the first practitioner-focused accounts of how information graphics are used and expected across disaster preparedness, warning dissemination, and response communication. 

\noindent\textit{C2: Identification of key challenges in disaster information graphics: }The study highlights common issues shaping the effectiveness of disaster graphics, including misinterpretation, information overload, inconsistent visual conventions, and limited accessibility guidance for complex formats such as hazard maps.

\noindent\textit{C3: Evidence of inclusiveness gaps affecting vulnerable populations: }We show how current disaster information graphics remain unevenly accessible for vulnerable communities, including people with vision impairments, cognitively impaired individuals, older adults, culturally and linguistically diverse groups, and people with low digital literacy.

\noindent\textit{C4: Recommendations and future directions for inclusive disaster graphics: }The study provides a way forward for disaster agencies, graphic designers, and researchers by outlining priorities for improving inclusive practice and motivating the need for adaptive frameworks to support accessible disaster information graphics.

The remainder of this paper is organised as follows. Section~2 reviews related work on disaster risk communication, information graphics, and inclusiveness challenges. Section~3 describes the study methodology. Section~4 presents findings from practitioner interviews, structured around the role of information graphics, current issues, and inclusivity concerns for vulnerable communities. Section~5 discusses the implications of these findings and outlines key directions for adaptive and inclusive disaster information graphics. Finally, Section~6 concludes the paper and highlights opportunities for future work.

\section{Background}

\subsection{Graphics in Communication}
Graphics have been used as a basic form of communication since the earliest stages of human history. Long before the emergence of standardised written languages, early societies, such as prehistoric cave dwellers, used cave paintings to record events \cite{valladas2001evolution}. These early graphics show that graphic expression emerged as one of the earliest structured communication mechanisms, which enabled humans to convey experiences and instructions even in the absence of formal text.

Information graphics are widely adopted in communication because of their unique cognitive and interpretive advantages. Firstly, \textit{graphics support rapid understanding than text}. Research in visual literacy suggests that humans are fundamentally visual learners, as visual information is processed by the brain 60,000 times faster than text \cite{burmark2002visual}. As a result, well-designed graphics can communicate relationships and key messages at a glance. This explains why diagrams, icons, maps, and infographics are frequently used in settings where information must be interpreted quickly, without requiring audiences to read lengthy explanations.
Secondly, graphics \textit{help to reduce cognitive load} by making complex relationships explicitly visible. Graphics can visualise complex information, so readers do not have to mentally construct meaning from the text. In the field of visual communication, this is known as externalising mental load \cite{josephson2020handbook}. E.g using line charts to represent trends over time, maps to show spatial structures in an area, step-by-step guides to show sequences of actions, and before-and-after figures to show comparisons. In doing this, graphics lower cognitive load and make information easier to interpret, particularly in situations where audiences may have limited time or attention. 
Thirdly, graphics also \textit{strengthen recall of information}. Visual information is often easier to remember than text. People tend to retain key messages better when they are supported by relevant pictures, diagrams, or structured visuals \cite{josephson2020handbook,lipkus1999visual}. This is explained through the principle of ``dual coding", whereby information presented through both verbal and visual channels is more likely to be remembered \cite{josephson2020handbook}. Such effects help explain why graphics remain central in educational, instructional, and public communication materials. 
Fourthly, \textit{visual elements attract attention and engagement}. Visual cues such as images, icons, and colour cues are highly salient and can draw audiences toward information that might otherwise be ignored \cite{bica2017visual}. For this reason, visual communication plays a central role in public-facing domains such as education, advertising, and mass media.

Despite these advantages, graphics also introduce significant limitations. One key concern is that \textit{graphics do not simply ``display" information; instead, they actively frame meaning}. Visual design choices such as colour, scale, layout and symbols influence what people perceive as important, urgent, or credible \cite{lipkus1999visual}. For example, if an infographic shows numbers, one number is highlighted in red and another in grey: Even if both values are equally important, readers will immediately interpret the red value as more urgent, problematic, or critical. Therefore, graphic designers need to pay attention to these aspects to avoid misleading the audience. 
Another limitation is that, unlike the common assumption, \textit{graphics are not universally understood}. Comprehension depends on conventions, education, cultural familiarity, prior experience, and accessibility needs \cite{lipkus1999visual,josephson2020handbook}. Charts, maps, and symbolic systems require visual literacy, meaning that what feels intuitive to one group may be confusing to another. The Handbook of Visual Communication reinforces the idea that visual meaning is socially learned rather than automatic, and that interpretation is shaped by context and shared conventions \cite{josephson2020handbook}. An article from National Geographic also discusses the cultural dependence of graphics in their article titled "Even graphics can speak with a foreign accent" \cite{EvenGrap68_online}. 
A final limitation of graphics is that \textit{poorly designed graphics can mislead or exclude}. Overly complex, ambiguous, culturally inappropriate, or inaccessible visuals may introduce misunderstanding rather than clarity \cite{lipkus1999visual,cyclonesconesconfusion}. For example, consider a public information poster that combines: multiple icons, several colours, dense text and a complex chart. Even though the goal here is to inform, audiences may not know where to look first or what the main message is. Such graphics may overwhelm audiences rather than clarify information, leading to confusion or disengagement. These limitations show that while graphics offer powerful communicative advantages, their effectiveness depends critically on careful design and an awareness of the audience and the context in which they are viewed, to help those who must make sense of them.

\subsection{Disaster Risk Communication}
Communication happens across many domains, such as health, education, and public information. Among these, disaster risk communication (DRC) is a particularly high-stakes context, where messages must be interpreted quickly and acted upon under stress and uncertainty. DRC refers to the processes through which information about hazards and protective actions is shared between authorities and the public. It is widely recognised as a central component of disaster risk reduction, supporting communities in understanding evolving threats and making timely decisions to protect the public \cite{bradley2016effectiveness}. International frameworks such as the Sendai Framework for Disaster Risk Reduction emphasise that clear, timely, and people-centred DRC is essential to reduce loss of life and harm during disasters \cite{SendaiFr7_online}. DRC is therefore not an optional activity, but part of the core infrastructure through which disaster resilience is built. DRC is not limited to real-time warnings; instead, it spans the entire disaster lifecycle, including preparedness, response, and recovery. In the preparedness phase, communication focuses on education and awareness-building; during emergencies, on alerts and situational updates; and after events, on recovery information and ongoing support \cite{stewart2024advancing}. The objectives of communication also shift across these phases, from building general understanding and readiness, to enabling rapid protective action, to supporting longer-term recovery processes. This temporal complexity means that DRC must adapt to changing contexts, audiences, and operational demands.

DRC has several features that differentiate it from other communication areas. A key characteristic of disaster risk communication is that it is \textbf{inherently action-oriented} \cite{stewart2024advancing}. Unlike general public information, risk messages are intended to influence protective behaviour under conditions of urgency and uncertainty. Effective communication is therefore measured not only by whether information is delivered, but by whether communities understand the message, recognise its relevance, and are able to act appropriately. Prior research highlights that communication failures can occur even when information is technically accurate, if it does not translate into meaningful decision-making or timely response \cite{liu_is_2017}. Another characteristic is that DRC is also \textbf{shaped by unavoidable uncertainty}. Hazards often evolve rapidly, and therefore, communicators must at times provide guidance with incomplete or changing information \cite{noji1997nature}. How uncertainty is communicated can strongly affect public interpretation and compliance with recommended actions during a disaster \cite{briere2000prevalence}. Trust is, therefore, central to DRC effectiveness. Communities are more likely to act on warnings when the source is perceived as trustworthy. Conversely, conflicting messages, misinformation, or past institutional failures can undermine confidence and reduce adherence to guidance. DRC is also \textbf{socially and culturally embedded}. The interpretation of risk information is shaped by community experience, cultural norms, language, and local knowledge \cite{briere2000prevalence}. As a result, risk communication cannot be assumed to operate uniformly across populations or settings. Current advances in DRC increasingly emphasise the importance of context-sensitive communication practices that account for how meaning is negotiated within specific communities \cite{briere2000prevalence,stewart2024advancing}.

In recent years, similar to other communication areas, a shift has occurred in DRC toward an increasingly multi-channel, digitally mediated approach \cite{niyazi2023application}. Communication now occurs through emergency apps, websites, SMS alerts, broadcast media, and social media platforms, often simultaneously. While digital systems enable rapid dissemination and real-time updates, they also introduce challenges such as information overload, platform fragmentation, and the rapid spread of rumours or misinformation \cite{talley2019disaster}. Disaster communication agencies must therefore navigate complex communication ecologies, where communities may receive multiple, and sometimes conflicting, streams of information \cite{hanspal2024role,bao2022use}. Recently, another change that has occurred in DRC is the growing recognition that it is most effective when DRC incorporates community engagement and feedback, rather than operating solely as a one-way transmission of warnings \cite{briere2000prevalence}. People-centred models of early warning systems highlight the importance of two-way communication, local intermediaries, and participatory approaches that build shared understanding before crises occur \cite{blake_get_nodate}. Such perspectives position DRC as an ongoing social process, rather than a purely technical delivery of alerts.

Overall, prior work establishes DRC as a critical, complex, and socially grounded component of risk reduction practice. It must operate across temporal phases, under uncertainty, within culturally situated contexts, and through increasingly digital communication infrastructures. These broader challenges provide an important foundation for examining how specific communication artefacts, including information graphics, function within disaster communication systems.

\subsection{Graphics in Disaster Risk Communication}

Graphics are widely adopted in DRC due to their ability to translate complex, technical information into forms that are quickly understood and easily remembered under stress. Prior work shows that infographics offer a concise and effective way to transfer critical information during emergencies, enabling rapid information sharing when time is limited \cite{lee2024visually, dootson2021managing}. Yavar et al. show particular value in the first 72 hours after a disaster, when stress impedes cognitive processing \cite{yavar_effective_2012}. Therefore, graphics, with their ability to minimise cognitive load and improve recall, are seen as an excellent communication format that encourages communities to take more decisive actions in DRC. Blake et al. also emphasise that pictorial resources can empower people to understand and independently apply preparedness actions, especially for vulnerable groups such as older adults, people with mobility challenges, and those with limited literacy \cite{blake_get_nodate}.

In DRC, maps play a significant role by conveying where the hazards are, who is at risk and where to go. They help people quickly gain knowledge about hazard extent, evacuation routes and local terrain~\cite{kurowski2011assessment, liu_is_2017}. Liu et al. specifically found that adding maps to short emergency warnings improves how at-risk publics comprehend, interpret, and act on the messages. \cite{liu_is_2017}. They also report that maps and graphical warnings outperform text-only messages in shaping risk perception and behavioural intent, while also noting the persistent risks of misreading scale, colour and uncertainty—emphasising the need for clear legends and consistent symbology. Additionally, a preparation infographic designed by the New Zealand government further illustrates how standardisation and multi-modal dissemination (print, online and social media) can improve public preparedness, with visual consistency building trust and recognition across agencies \cite{blake_get_nodate}. In summary, graphics play an important role in DRC and help to connect citizens, emergency services, and policymakers through a shared visual vocabulary that can cut across literacy, language, and experience differences.

Despite clear benefits, the design and deployment of graphics in DRC present three persistent problems: a lack of user engagement, inadequate communication, and a lack of inclusiveness. The first of these, \textbf{absence of user engagement}, is discussed in several studies. These studies report limited adoption of user-centred design (UCD) and weak evaluation practices in disaster visualisation. Twomlow et al. note a systemic shortfall in validation and user testing, with few projects implementing a full UCD cycle or incorporating vulnerability and social data \cite{twomlow_user-centred_2022}. They argue for co-design and iterative testing with end-users to prevent design bias and miscommunication, and emphasise that colour/icon meanings are culturally contingent—e.g., red can signal danger in some contexts but joy in others—so contextual adaptation is essential. The second problem noted in the literature is \textbf{conflict between Aesthetics and Cognitive load}. Visuals often privilege aesthetics or density over clarity. Yavar et al. caution that attractive design should not override informational accuracy; highly decorative or overloaded graphics increase cognitive load in time-critical situations \cite{yavar_effective_2012}. Experimental work also shows that poor text–graphic integration undermines comprehension: pictures unrelated or loosely linked to text do not aid understanding and can distract viewers \cite{houts_role_2006}. In the hazard-mapping space, adding maps to alerts can modestly improve decision clarity, but misreading of scale, colour and uncertainty remains common—underscoring the need for clear legends and consistent symbology \cite{liu_is_2017}. The last problem that is discussed is the \textbf{lack of inclusiveness}. Access to risk information is uneven across disability, literacy, language and socioeconomic lines. As this is a major focus of our research as well, we talk about this problem in detail in the next section.




\subsection{Status of Inclusivity in Disaster Risk Communication}
Although disaster risk communication research increasingly recognises inclusiveness as a core concern, the specific inclusivity of disaster information graphics remains comparatively under-examined. Much of the existing literature focuses on improving the inclusivity of DRC more broadly, providing limited detailed analysis of how visual artefacts such as maps, warning icons, infographics, and evacuation diagrams may exclude or disadvantage vulnerable audiences. Few studies have attempted to improve the inclusivity of graphics in DRC, and in this section we summarise these works. We also draw from other fields, such as health communication, to understand their initiatives to improve graphic inclusivity.  

A major concern for uneven accessibility is related to people with \textbf{disabilities}, particularly those with visual, auditory, or cognitive impairments. Twomlow et al. is one of the few who have studied this from a visual DRC perspective, and they emphasise that inclusive disaster visualisations require adherence to established accessibility guidance (e.g., W3C standards), proactive testing with assistive technologies, and the development of simplified or alternative formats \cite{twomlow_user-centred_2022}. They further note that many commonly used digital hazard systems and tools remain incompatible with screen readers, highlighting persistent barriers in practice. New Zealand's `Get Prepared' infographic project is one running example we will refer throughout this section as they have sought to increase inclusivity of DRC graphics. This study operationalises inclusive design principles through high-contrast palettes, large fonts, and descriptive alternatives, developed in collaboration with disability organisations such as Deaf Aotearoa \cite{blake_get_nodate}. 
They have also explored inclusivity from an \textbf{age} perspective and have explicitly incorporated older adults into co-design, adapting the content and presentation with simplified instructions, larger font sizes, and clearer iconography. 

Inclusiveness also depends strongly on \textbf{localisation} and place-based context. Effective disaster graphics must reflect local hazard conditions, infrastructures, and everyday practices rather than assuming universal transferability \cite{van2017understanding}. Twomlow et al.’s user-centred design framework embeds contextual grounding through iterative cycles that require early identification of user characteristics and hazard environments before visual encodings are finalised \cite{twomlow_user-centred_2022}. Related work on digital vulnerability further demonstrates that access to risk information varies significantly across locations and socio-technical settings. Hao et al.\ highlight that residents in subsidised housing may face compounded barriers to digital-first warning systems, reinforcing the need for localised and multimodal communication strategies that supplement online dashboards with print, radio, and community-based channels \cite{hao_examining_2022}. Earlier studies similarly argue that combining static and dynamic visual formats across multiple media remains essential for reaching diverse local audiences \cite{yavar_effective_2012}. 

\textbf{Language} accessibility is closely intertwined with localisation, particularly for culturally and linguistically diverse communities. Multilingual access is often recognised as important, yet inconsistently implemented in practice \cite{uekusa2023preparing}. The `Get Prepared' materials provide multilingual versions and plain-language phrasing as part of inclusive preparedness communication \cite{blake_get_nodate}. Hao et al. similarly show that non-English speakers are disproportionately excluded by English-only, digital-first risk information, and advocate for multilingual and low-bandwidth alternatives \cite{hao_examining_2022}. Supporting evidence from visual communication research suggests that pairing brief text with clear pictograms can improve comprehension when language proficiency is limited \cite{houts_role_2006}. 

Finally, \textbf{cultural context} plays a critical role in determining whether disaster graphics communicate meaning effectively. Visual conventions are not universal: colours, symbols, and metaphors may carry different meanings across societies \cite{kulatunga2010impact, maldonado2016considering}. Twomlow et al. caution that design assumptions about visual signals (e.g., the use of red to connote danger) may not be uniformly adopted, making co-design and contextual adaptation essential \cite{twomlow_user-centred_2022}. Houts et al. similarly recommend culturally familiar imagery to enhance attention and trust \cite{houts_role_2006}. The `Get Prepared' project demonstrates this cultural tailoring through partnerships with community groups to ensure that preparedness visuals align with local norms and practices \cite{blake_get_nodate}.

Overall, this prior research highlights that inclusivity in disaster information graphics is shaped by multiple intersecting factors rather than a single design adjustment. While emerging frameworks and inclusive preparedness initiatives provide promising direction for the inclusion of DRC as a whole, there remains limited systematic guidance for producing disaster graphics that are accessible, culturally appropriate, and actionable across diverse community needs.

\subsection{Summary}
In summary, prior work establishes that information graphics play a central role in communication because they support rapid understanding, reduce cognitive burden, and improve recall. Disaster risk communication represents one of the most critical contexts in which these benefits matter, as messages must often be interpreted and acted upon under stress, uncertainty, and time pressure. Accordingly, graphics such as hazard maps, warning icons, evacuation diagrams, and preparedness infographics have become widely embedded in contemporary disaster communication infrastructures.

At the same time, the literature also highlights persistent challenges. Graphics in disaster contexts can introduce misinterpretation, cognitive overload, and inconsistent visual conventions, particularly when they are designed without sustained engagement with end users. Most importantly, while inclusiveness is increasingly recognised as a key concern in disaster risk communication more broadly, the inclusivity of disaster information graphics remains comparatively underexamined. Existing work points to important dimensions shaping accessibility and interpretation, such as disability, age, language, localisation, and cultural context — but a systematic empirical understanding of how practitioners perceive these challenges and how inclusivity is currently addressed in practice remains limited.

These gaps highlight the need for a closer examination of disaster information graphics as real-world communication artefacts: how they are used, the issues that emerge in practice, and their inclusivity for vulnerable populations. To address this need, we conducted a qualitative interview study with disaster communication practitioners and researchers. The next section describes our study design, participant recruitment, data collection process, and thematic analysis approach.

\section{Methodology}

This study investigates how information graphics are used in DRC and how inclusive these graphics are for vulnerable populations. Given the exploratory and practice-oriented nature of the research questions, we adopted a qualitative approach, using semi-structured interviews with disaster communication practitioners and researchers.

We conducted a qualitative interview study to understand current uses, challenges, and concerns about inclusiveness related to disaster information graphics. Semi-structured interviews were chosen to allow participants to describe their experiences in depth while still covering a consistent set of topics across interviews. This approach is commonly used in disaster risk communication research to capture practitioner perspectives and operational realities.

\subsection{Participants}
Five participants (P1--P5) were recruited for the study. Participants included practitioners and experts working in disaster communication, emergency management, and disaster-related research. They represented both academic and emergency agency perspectives and had experience designing, using, or evaluating information graphics in disaster contexts. Participants were selected using purposive sampling to capture expertise relevant to disaster warning communication and the use of visual materials such as maps, action guides, and public warning graphics.

\subsection{Data collection}
Data were collected through semi-structured online interviews. Each interview lasted approximately 45- 60 minutes. Before the interview, participants provided basic demographic information, including their professional role, organisational context, and years of experience in disaster management. The interview questions were informed by an initial review of prior research on disaster risk communication, information graphics, and accessibility challenges in warning systems. This ensured that the protocol covered established themes in the literature while also allowing participants to raise new issues grounded in practice. 

The interview questions focused on three main areas: 1) Participants’ understanding of what information graphics mean in disaster communication, 2) Current status of graphics and understanding of any issues, 3) Inclusiveness challenges for vulnerable communities. Participants were also asked whether they had observed ineffective or misleading graphics, how graphics were currently designed in their organisations, and what limitations exist in making disaster graphics accessible in practice. All interviews were audio-recorded with consent and transcribed for analysis. Quotes included in the paper are attributed using participant identifiers (P1--P5).

\subsection{Data analysis}
Interview transcripts were analysed using reflexive thematic analysis. We followed an iterative coding process to identify themes related to:
\begin{itemize}
    \item The role of information graphics in disaster communication
    \item Issues in the design and interpretation of disaster graphics
    \item Inclusiveness gaps affecting vulnerable communities
\end{itemize}

Initial codes were developed through close reading of the transcripts. These codes were then grouped into broader themes and refined through repeated comparison across participants. The analysis aimed to preserve the practitioner-oriented nature of the findings, with themes grounded in participant language and supported through direct quotations.

\textbf{Ethical considerations: }The study was conducted in accordance with institutional ethics requirements. All participants provided informed consent prior to participation. Participation was voluntary, and transcripts were de-identified to protect confidentiality.

\textbf{Trustworthiness and rigour: } Several strategies were used to strengthen the trustworthiness of the analysis. Interviews included participants from both academic and operational emergency management contexts, allowing triangulation across perspectives. The use of direct quotations provides transparency in how themes were grounded in participant accounts. Coding was conducted iteratively, with themes refined to ensure consistency across the dataset. Although the sample size was small, participants provided rich insights into current practices and limitations in inclusive disaster information graphics, thereby supporting the exploratory goals of this study.

\section{Results}

\subsection{Demographics}

We interviewed five experts involved in disaster communication, representing both academic and operational perspectives as shown in Figure \ref{tab:participants}. This included three researchers from the UK and Australia with 8–10 years of experience, and two senior practitioners working in public information roles within Australian emergency-management agencies, with 12–16 years of experience. These participants provided valuable insights into how information graphics are used in both research and real-world disaster communication practice.


\captionof{table}{Participant Demographics and Expertise}
\label{tab:participants}
\begin{tabularx}{\textwidth}{lllll}
\toprule
\textbf{ID} & \textbf{Sector} & \textbf{Professional Role} & \textbf{Exp.} & \textbf{Region} \\
\midrule
P1 & Academic & Associate Professor & 10 yrs & UK \\
P2 & Academic & Professor & 8 yrs & AU \\
P3 & Academic & Assistant Professor & 8 yrs & UK \\
P4 & Industrial & Senior Advisor, Public Information & 12 yrs & AU \\
P5 & Industrial & Manager, Public Information & 16 yrs & AU \\
\bottomrule
\end{tabularx}

\subsection{Use of Graphics in Disaster Communication}

\subsubsection{Different Types of Graphics} 

\begin{figure}
    \centering
\includegraphics[width=\textwidth, trim={0cm 5cm 0cm 0cm},clip]{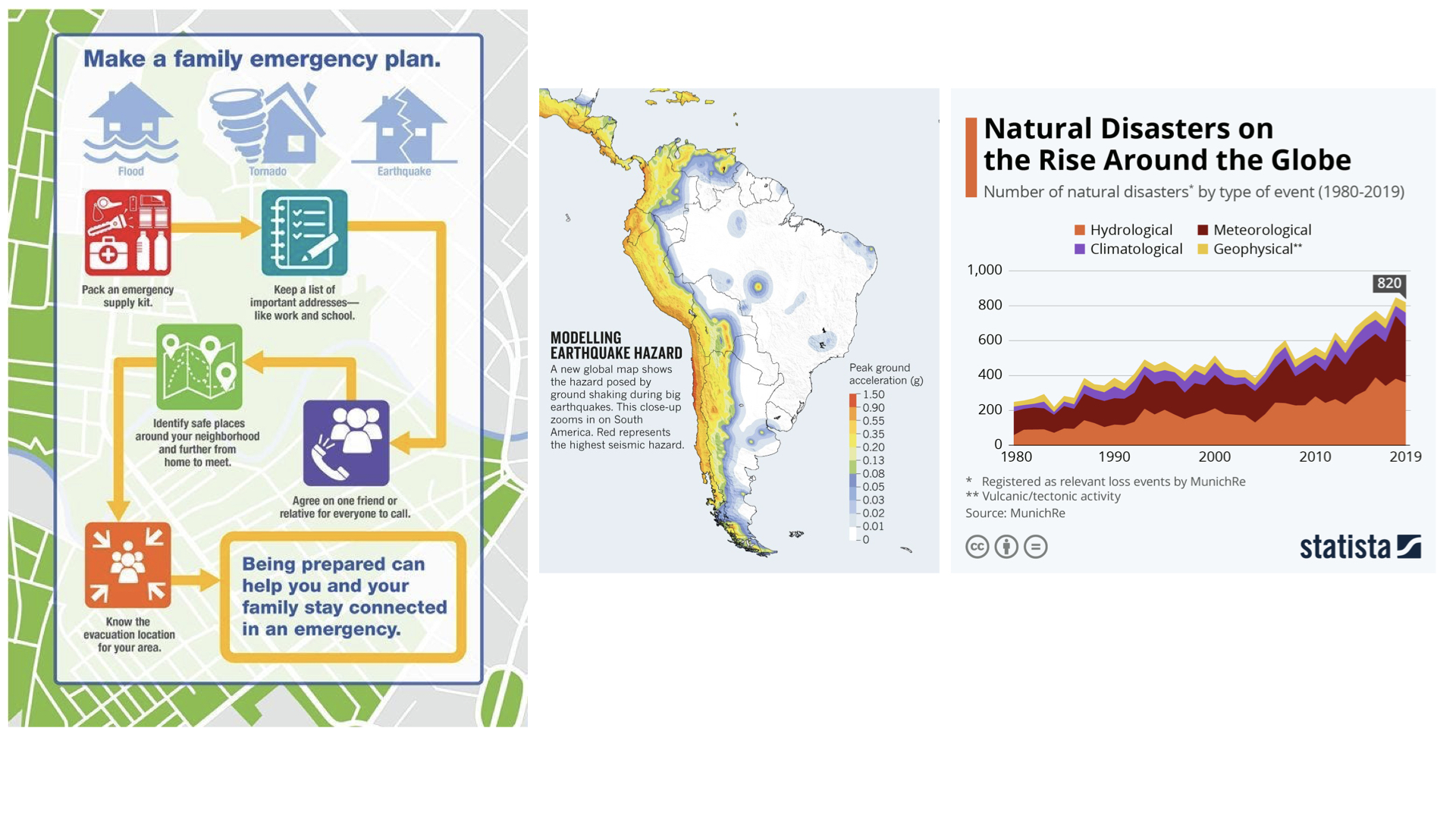}
    \caption{Sample Types of Graphics (a)Pictorial Action Guide \cite{HowtoCre12_online}, (b)Hazard Map \cite{Earthqua36_online}, (c)Statistical Chart \cite{ChartNat48_online}}
    \label{fig_sample}
\end{figure}

Participants described information graphics as an integral and widely used component of disaster communication, used throughout the disaster lifecycle, including preparedness campaigns, warning dissemination, and response coordination. A wide variety of information graphics were discussed, depending on the hazard type, the urgency of the message, and the intended audience. These included pictorial action guides, maps, statistical charts and many more. Samples of some of these are shown in Figure \ref{fig_sample}. Practitioners highlighted how graphics are often used not only to issue warnings but also to explain risk in ways that feel concrete and relatable. As one participant explained, they use visuals to \textit{“provide warnings to the community for those [flood, storm, earthquake, tsunami, and landslide] hazards, and also how we actually explain the risk to people ... how do we show the risk of driving a car through floodwater, for example, using a diagram”} (P4).

Participants also emphasised the instructional role of disaster graphics in guiding immediate action. For example, one academic described how, in parts of Indonesia, there is a protocol that each meeting should begin with graphics of visual drills or short videos that show \textit{“If there is a disaster, what you have to do, where is the exit point”} (P2). Such examples illustrate how information graphics are used not only for awareness, but also for reinforcing practical response behaviour.

\subsubsection{Media of Graphic Presentation}

Participants discussed how information graphics are delivered in both print and digital formats. Examples for printed materials included posters and leaflets, while digital formats included graphics shared through websites, mobile phone apps, social media, public display screens (e.g., elevators and community centres), and television broadcasts.

Both academic and emergency-agency participants noted a clear shift toward digital-first communication. As P3 explained, \textit{“More than 90\% of them [graphics] are digital. It used to be different — in the past, we’d draw those things physically, with paper and pen, but that stopped long ago. Everything we create now — even flowcharts — is digital”} (P3). Practitioners also described how digital workflows support rapid adaptation and redistribution across channels, including print. One participant noted that agencies often prioritise producing \textit{“a digital product that’s easily printed”} (P4). This highlights how disaster information graphics are increasingly created for flexible multi-platform use, allowing the same visual message to circulate quickly across both online and physical environments, with digital being the most used format.

\subsubsection{Purpose of Graphics}
Our participants emphasised that information graphics are central to how disaster risk information is communicated in practice. They described graphics as a key mechanism for translating risk knowledge into clear, actionable guidance for communities. Visual formats were seen as particularly important during disaster situations, where stress, uncertainty, and time pressure can limit the public’s ability to engage with dense textual information. This was emphasised by P2 when they talked about how Non-Government Organisations use graphics for disaster communication: \textit{``it’s [graphics] easier to communicate to the general public. They understand better compared to if you put it in text."} (P2)

Practitioners also noted that emergency agencies increasingly prioritise producing graphics that ``hit the mark'' (P3) in terms of urgency and clarity, especially in fast-moving hazards such as bushfires and floods. In this sense, graphics were not viewed merely as explanatory artefacts, but as communication tools intended to prompt timely protective action.

\subsubsection{Intended Audiences and Effectiveness of Graphics}

Participants explained that most disaster information graphics are typically designed for broad communication with the general public. Few discussed graphics designed specifically for NGOs, decision-makers, academics, civil protection authorities, and municipalities. 
However, several participants highlighted that different forms of graphics are suited to different audiences. For example, maps were often viewed as more appropriate for visually oriented users and younger community members who are accustomed to navigation tools on mobile devices. In contrast, condensed statistical figures and summary charts were seen as particularly useful for communicating with decision-makers who require rapid situational understanding. As one participant explained, such formats are valuable when presenting information to individuals who may not be familiar with disaster management but still need to make quick, high-level decisions: \textit{``They want condensed information: accessible, visual, quick to grasp''} (P3). 

Our participants also discussed several strategies currently used in practice to increase the effectiveness of these graphics. One common approach is to place highly visible, large-format visuals in prominent community locations. For example, one participant described how preparedness posters were displayed in central public spaces: \textit{``On the board in front of the village head office... every day they will see that. So it's kind of embedded in their mind''} (P2). Several participants similarly highlighted the use of large graphics, both physical and digital, in high-traffic environments such as elevators, community centres, and other shared public spaces.

Another strategy emphasised the value of simple, step-based pictorial guidance that can be understood even when textual information is inaccessible. One participant described being able to follow earthquake preparedness visuals despite not understanding the accompanying language, noting that \textit{``it’s very easy to see the steps''} (P2).

The practitioners also noted the growing use of engagement metrics to assess communication reach. One emergency management participant described how agencies rely on online statistics to evaluate visibility and uptake: \textit{``We use numbers like hits into the website... how many people viewed it, how long they're on that page''} (P5).


Together, these findings highlight that while disaster information graphics are typically developed for broad dissemination, practitioners recognise the need for more audience-sensitive approaches and clearer strategies to ensure effectiveness across diverse communities.

\subsubsection{Designing graphics}

Participants described the design of disaster information graphics as a practical, workflow-driven activity, rather than something owned by a single ``designer.'' Instead, graphics are often produced through a combination of in-house communication teams, operational staff, and external collaborators, depending on organisational capacity and context.

Several practitioners noted that emergency agencies commonly develop graphics internally, often drawing on established templates and shared communication processes. One participant explained that these materials are usually created by \textit{``our in-house team... combining staff from multiple agencies''} (P5). They also highlighted that agencies rely heavily on reusable formats: \textit{``In terms of visual products, we have many templates ... We try not to create things from scratch on the fly,  everything is pre-set to make it quick''} (P5).

At the same time, participants emphasised that dedicated design expertise is not always widely available. One practitioner noted that while agencies may employ a graphic designer, capacity is often limited: \textit{``We have an in-house graphic designer, but we have one to service the entire state [in Australia]''} (P4). Outsourcing was described as possible only in certain funded situations, where \textit{``particular projects may have money to outsource and actually provide professional support''} (P4).

Participants also pointed out that not all disaster information graphics originate within emergency agencies. Some visuals are developed directly by researchers or scientists, which can create additional challenges when translating technical information for public communication. As one academic noted, \textit{``One of the examples I gave was designed by scientists, not agencies''} (P1).

Overall, participants framed graphic design as a collaborative but sometimes fragmented process. Rather than a single end-to-end pipeline, graphics are shaped by multiple actors contributing different forms of expertise: \textit{``.. a lot of the time it's just a fair bit of us doing our best with the support of our media and communication team''} (P4).


In this sense, the design of disaster information graphics is closely tied to institutional constraints, time pressure, rather than being a purely technical or aesthetic design task.

\subsection{Issues in Disaster Graphics}

Participants identified a range of recurring challenges in the current use of information graphics in disaster communication. While these graphics are widely relied upon for public warnings and preparedness messaging, experts consistently emphasised that their effectiveness is not universal. Significant barriers remain in terms of how graphics are interpreted, who they reach, and how they are produced within organisational practice. A summary of these issues is shown in Figure \ref{issues}

\begin{figure} 
  \includegraphics[width=\textwidth, trim={0cm 20cm 16cm 0cm},clip]{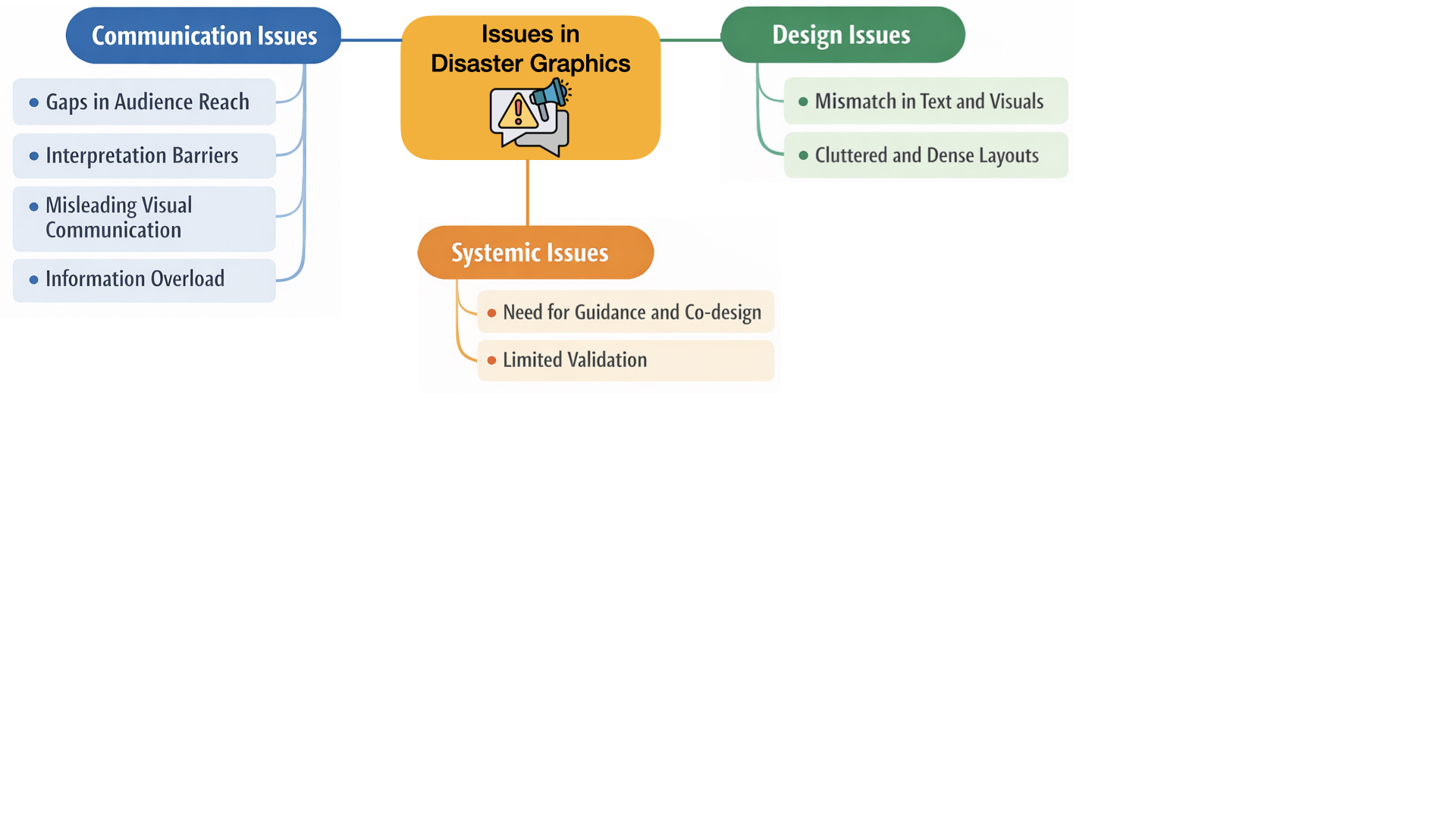}
  \caption{Overview of Issues in Disaster Communication Graphics}
  \label{issues}
\end{figure}

\subsubsection{Communication Issues}

\textbf{Gaps in Audience Reach: }
A key concern raised by participants was that disaster information graphics are often developed to communicate with the broad majority, rather than being designed with specific community needs in mind. One practitioner reflected that disaster communication is frequently \textit{``just about reaching the majority and not always as targeted as going into particular groups''} (P4). As a result, marginalised populations may be less likely to receive or interpret warning information effectively.

Participants also highlighted that certain groups may struggle with more technical or data-driven visual formats. For example, one participant observed that 
\textit{``those without a scientific or formal education background... are less familiar with data or charts''} (P1), while another similarly pointed out that 
\textit{``non-specialists unfamiliar with reading scientific or spatial information''} (P2) may find maps or data visualisations difficult to interpret. P1 reinforced this with an example from their own work, noting that community members 
\textit{``didn’t read bar charts — they couldn’t understand them... we needed alternative approaches, like using icons or symbols, or adding more explanation.''}

\textbf{Interpretation Barriers: }
Participants noted that disaster graphics often assume a baseline level of visual and technical literacy that may not hold across communities. Even when visuals are designed to simplify information, their meaning can remain ambiguous or dependent on prior familiarity, as expressed by P3, some people struggle with spatial representation and \textit{''some groups are less map-literate"} (P2). This creates risks of misunderstanding, particularly in disaster situations where stress, urgency, and limited attention already reduce cognitive capacity. Overall, experts emphasised that information graphics cannot be treated as automatically self-explanatory artefacts. Interpretation varies substantially depending on audience background, experience, and situational context.

\textbf{Misleading Visual Communication: }
Participants also raised concerns that graphics can be strategically manipulated, intentionally or otherwise, leading to misleading representations. As one participant cautioned, \textit{``it’s very easy to deceive people with graphs or maps... it’s easy to manipulate audiences based on how you present data''} (P3). Expanding on this, P3 gave an example of how graphs can be used to shape public perception and resource allocation: \textit{``I’ve seen graphs used to exaggerate issues — for example, representing drought-affected areas in a way that makes the problem look worse than it is, to attract more resources,”} and they added that this may be \textit{“sometimes intentional, sometimes due to ignorance or poor communication.”} (P3). Another example given was ``Mercator map, for instance — it’s been used forever, placing Europe in the centre, making Greenland larger, Africa smaller — giving implicit bias and importance to certain regions'' (P3). This shows that miscommunication may not always be deliberate, but the consequences, distorted understanding and mistrust remain significant.

\textbf{Information Overload: }
Participants also described a growing expectation that disaster information should be centralised and integrated, particularly through map-based interfaces. P4 reported that \textit{``people want everything in the one spot,''} and that they want to be able to \textit{``go to a map to see all the information.''} However, P4 also highlighted an important tension: attempts to consolidate information are often 
\textit{``not hitting the mark''} because the result becomes \textit{``too confusing and not precise enough for what people want''}. This suggests a trade-off between completeness and clarity, where ``all-in-one'' visual communication can unintentionally reduce the understandability of graphics.

\subsubsection{Design Issues}

\textbf{Mismatch in Text and Visuals: }
Participants highlighted that text and visuals are not always integrated in ways that support comprehension. One observed that communicators often attempt to create images with short accompanying text, but that \textit{``some of the short facts are a bit funny...''} {P2)}. This implies that poorly aligned or unclear textual content can dilute the intended message and increase the effort required for interpretation.

\textbf{Cluttered and Dense Layouts: }
Another recurring design problem was overloaded layouts that reduce readability, especially during multi-hazard events. P4 described ``real issues'' when multiple hazards must be shown at once in real-time digital maps: \textit{``We could have storms, floods and fires potentially all occurring on the same day. So we're actually layering colors over colors and layering information over information, with icons thrown into the mix as well, where information is just getting lost.''}. This indicates that current graphic systems can struggle to maintain hierarchy and clarity as information density increases, precisely when communities need the clearest possible guidance.

\subsubsection{Systemic Issues}

\textbf{Need for Guidance and Co-design: }
Participants described a systemic gap in user-centred and co-designed processes for producing disaster information graphics. P1 argued that \textit{``user-centred design would be important''} to better consider community needs and communicate risk more effectively. In practice, however, practitioners noted limited pathways and expertise to support inclusive design. P4 suggested that organisations such as Vision Australia and accessibility guidelines could provide a starting point, but acknowledged that \textit{``I probably haven’t been involved in trying to sort of get to that level.''}. This need for structured support was echoed at the organisational level. P5 stated: \textit{``We’d love further guidance,''} describing interest in funded partnerships focused on \textit{``co-design... multilingual, multicultural communities... co-designing to create warnings and other products.''}

\textbf{Workflow Constraints: }
Participants also highlighted systemic barriers related to who produces disaster information graphics and the constraints under which they are produced. Practitioners noted that graphics are often developed within institutional teams under significant time pressure, particularly during emergencies. Responsibility for design is fragmented across departments or outsourced, and inclusiveness considerations are not always embedded in standard workflows. \textit{``And a lot of it comes down to whether we have the capability with the time constraints we've got with our warnings product" }(P4). This suggests that several limitations of disaster graphics also reflect issues with organisational structures rather than isolated design oversights.

\textbf{Limited Validation}
Another systemic issue is the lack of validation of whether graphics are actually effective for public understanding and action. P1 noted that \textit{``we didn’t do any validation or effectiveness testing,''} even though they believed visuals were \textit{``a good way to communicate scientific data.''} Where evaluation does occur, it is often narrow or constrained. For example, P5 described testing that was \textit{``just workshop visual testing,''} rather than a more comprehensive assessment with diverse individuals and vulnerable groups.

Together, these accounts suggest that current production pipelines often prioritise rapid delivery over evidence of comprehension, usability, and inclusiveness.

\subsection{Inclusivity of Graphics for Vulnerable Populations}

When asked about inclusiveness in disaster communication, participants consistently mentioned a wide range of vulnerable communities that may not be equally represented in current disaster information graphics. These groups included disabled communities (vision-impaired, colour-impaired, cognitively impaired) as well as other communities such as elderly populations, children, local communities in hazard-prone areas, culturally and linguistically diverse groups, indigenous communities, women, and younger audiences. Below we describe the discussions on inclusivity related to these groups.

\subsubsection{Barriers Faced by Vulnerable Communities with Disaster Graphics}

\textbf{Vision Impaired Communities: }
Participants frequently identified vision-impaired communities as a key group that current disaster information graphics do not always support effectively. P2 emphasised that many disaster visuals remain strongly dependent on sight, noting that \textit{``all these graphics...it needs vision. So low vision, people will have difficulties...mostly there is no alternative besides vision''} (P2). This reflects how visual-first warning communication can exclude low-vision audiences when alternative formats are not consistently available.

Participants also highlighted that accessibility guidance for complex visual interfaces remains limited, particularly for map-based communication. P5 explained that accessibility frameworks often treat mapping as especially challenging, describing that \textit{``a lot of accessibility frameworks just say `Mapped things are too hard, so you just create a list instead''' } (P5). 

In response, practitioners described several design-level adjustments to improve visual accessibility. For example, P4 noted efforts to simplify colour use in communication materials, such as \textit{``removing colors so materials can be printed in black and white — using monochromatic type colors''} (P4). However, participants suggested that such approaches remain partial and that achieving consistent support for vision-impaired users remains difficult in practice.

\textbf{Cognitive Impaired Communities: }
Participants suggested that disaster information graphics can impose high cognitive demands, making them difficult to interpret, particularly under stress. P4 noted that \textit{``information is just getting lost for someone who doesn’t have any sort of impairment,''} adding that \textit{``it’s hard enough for me...to actually understand a lot of it,''} meaning that \textit{``your average person out in the community is probably going to struggle as well''} (P4). These challenges were described as becoming more pronounced when multiple hazards must be communicated at once. P4 explained that emergency graphics can involve \textit{``layering colors over colors and layering information over information, with icons thrown into the mix,''} which can undermine comprehension during time-critical situations.

Participants also noted limited attempts to reduce cognitive load through clearer, more sequential visual guidance. For example, P4 described the use of step-by-step pictorial formats, such as \textit{``very clear graphics that show step-by-step visuals...so you wouldn’t need to read the words''} (P4). However, such approaches were not described as consistently embedded within mainstream warning communication.

Cognitive accessibility issues were also evident in the design of complex visual interfaces. P5 explained that while users may want to \textit{``see the map [of hazards] and see the list [of hazards]...together,''} \textit{``in an accessible sense, that just doesn’t work''} (P5). As a result, systems may separate information into a \textit{``full map view''} and a \textit{``full list view''} (P5), requiring users to navigate different representations rather than receiving integrated situational information.

\textbf{Elderly Communities: }
Older adults were also repeatedly identified as a key group requiring better support. However, participants noted that disaster communication materials are rarely tailored specifically for elderly users. As P1 explained, \textit{``Specifically for elderly — probably no''} (P1).

\textbf{CALD Communities: }
Inclusiveness was also discussed in terms of cultural and linguistic fit. Participants highlighted culturally and linguistically diverse communities as an important priority in disaster communication, although progress was often described as limited. One approach currently taken is to simplify materials so they can be more easily adapted when needed. For example, P4 explained that agencies try to ensure that \textit{``we make products as plain English as possible so they’re more easily translated when required''} (P4). However, participants noted that further multilingual support remains constrained, particularly during real-time emergencies.

Practitioners also cautioned that disaster information graphics are often treated as transferable templates, even though safe actions may differ significantly across contexts. P2 highlighted how earthquake guidance may not translate safely across countries, explaining that \textit{``the way Japanese houses...are built is different...you probably shouldn’t hide under the table...(it’s) not safe in Indonesia''} (P2). This suggests that cultural and environmental differences can fundamentally shape whether graphic guidance is valid in local settings.

Participants further raised concerns about linguistic accuracy and cultural sensitivity in translated materials. P2 described bilingual resources where \textit{``some of the English are very bad, the way they translate,''} even if \textit{``the pictures are very clear''} (P2). Beyond language quality, P2 also questioned whether using the same personas across regions may be inappropriate, noting uncertainty about \textit{``how culturally sensitive that persona is...and whether...culturally insensitive...affect the effectiveness''} (P2).

\textbf{Young Communities: }
Some participants described efforts to tailor visuals to younger age cohorts, but these were not always central to core warning materials. P3 noted that \textit{``if we targeted younger people — we’d think about color and design''} (P3). Similarly, P5 noted that work aimed at younger audiences has occurred only as a secondary focus, stating that they \textit{``did do a small piece connected to Gen Z and very young audiences. But it was a secondary piece of work, a secondary target''} (P5)

\textbf{Gender- Women: }
P5 noted that gender-focused work has only recently begun, explaining that agencies have traditionally not focused specifically on women, but are now starting to explore this space. In particular, P5 described a recent campaign targeted directly at women aged 35–55, based on their caregiving and planning roles in emergencies and their high engagement with emergency communication platforms: \textit{``We ran a campaign targeted directly at women aged 35 to 55 because of the carer role that they undertake within emergencies. They are actually our most engaged community on VicEmergency.''} (P5).

\textbf{Education and Low Literacy: }
Participants linked inclusiveness to educational differences and the ability to interpret common data formats. P4 described efforts to provide \textit{``plain English versions,''} but acknowledged that this is constrained by resources: \textit{``it’s about what we can actually achieve with the budget we’ve got''} (P4). Likewise, P1’s pilot testing revealed that some users \textit{``didn’t read bar charts — they couldn’t understand them''} (P1).

Participants noted that simplifying graphics for low-literacy audiences can improve inclusiveness. P1 described shifting away from complex representations after testing. This emphasis on simplification was echoed in design choices aimed at reducing barriers: \textit{``We try to make things as simple as possible,''} and \textit{``making things as simple as we can''} (P1).

\textbf{Other: }
Participants also referred to children, women and local communities living in hazard-prone areas as groups that are often considered vulnerable but are not always explicitly addressed in targeted design.

\subsubsection{Current Practices in Addressing Vulnerability}

Participants described several ways in which agencies and researchers currently attempt to address vulnerability and inclusiveness in disaster information graphics. These practices include design-level adjustments, limited accessibility testing, reliance on community intermediaries, and operational strategies shaped by preparedness workflows and resourcing constraints. However, participants also emphasised that current efforts remain uneven and difficult to sustain, particularly during real-time emergencies.

\textbf{Design Adaptations and Simplification Efforts: }
One common approach involves simplifying visual formats in order to improve comprehension for broader audiences: \textit{``We try to make things as simple as possible"}(P4). Participants noted that complex charts and statistical representations are not always interpretable for all community members, particularly in high-stress contexts or for those with lower visual literacy. For example, P1 explained that pilot testing revealed some users \textit{``didn’t read bar charts — they couldn’t understand them,''} highlighting the need for more intuitive alternatives such as \textit{``icons or symbols, or adding more explanation''} (P1). These accounts suggest that current inclusiveness efforts often begin with reducing complexity and shifting toward more concrete, action-oriented visual encodings.
.

\textbf{Accessibility Testing and Challenges in Sustained Engagement: }
Participants also described attempts to improve inclusiveness through user testing with diverse groups. P3 noted that they have \textit{``tried to make them accessible for different vulnerable groups,''} and would sometimes \textit{``include diverse users in testing sessions, to check visibility and accessibility''} (P3). However, such engagement was often described as difficult to sustain. P5 reported that it was \textit{``very difficult to get engagement,''} and that disability-community pilots produced only \textit{``very minimal feedback''} (P5). In one instance, P5 noted that \textit{``we had two people from Scope and three, four people...and got very minimal feedback from them,''} suggesting that recruitment and iterative evaluation remain major practical constraints. Participants further acknowledged that inclusiveness efforts are \textit{``not always successful''} (P3), reflecting the challenges of ensuring representative participation across different vulnerable communities.

\textbf{Reliance on Local Intermediaries and Community Networks: }
Rather than systematic workflows, participants explained that agencies often rely on local intermediaries and community leaders to understand vulnerability-related needs, particularly in multicultural contexts. P4 described how they \textit{``tap into our local resources...to understand what the needs of the community are,''} including whether communities \textit{``may need to have translated materials''} or require formats that are \textit{``not written materials''} (P4). Participants also linked inclusiveness to representation and legitimacy. P3 noted that \textit{``communities and NGOs...want to see themselves and their contributions represented''} (P3), suggesting that inclusive disaster graphics involve not only accessibility, but also whose perspectives are made visible within official communication materials.

\textbf{Preparedness versus Response-Phase Constraints: }
Participants emphasised that inclusive practices are often more feasible during preparedness activities than during real-time emergency response. While preparedness materials allow time for consultation and adaptation, response-phase communication is shaped by urgency, workload, and limited operational capacity. As P4 explained, agencies are \textit{``better at doing it ahead of an emergency...to ensure that we can reach multiple groups,''} but during the actual response phase \textit{``we’re probably not as able, at the moment, to do some of that stuff''} (P4). This was particularly evident for multilingual accessibility. P4 noted that \textit{``a lot of it comes down to whether we have the capability,''} and that while agencies may prepare translated materials in advance, \textit{``during the actual response phase...we’re probably not as able''} (P4). These accounts highlight how inclusiveness is shaped by operational realities, not solely by design intentions.

\textbf{Funding and Resource Limitations: }
Finally, participants repeatedly pointed to resourcing constraints as a key limitation in sustaining accessibility work. P5 emphasised that beyond an initial pilot, \textit{``we haven’t had any specific funding...to really create anything else,''}. The same participant further mentioned that \textit{``We’ve been focusing for the last five years on improving accessibility and multilingual reach, though funding limits progress"} (P5). These show that accessibility efforts often depend on \textit{``best efforts''} rather than institutionalised support (P5). Translation and alternative formats were also described as reactive rather than embedded. P5 noted that during an emergency, \textit{``the money sort of just appeared''} (P5), indicating that accessibility provisions may only become possible in urgent situations rather than through sustained investment.

Overall, participants described current practices as involving partial design adaptations, limited evaluation with vulnerable users, and ad hoc reliance on community networks. However, these efforts remain constrained by engagement challenges, the urgency of the response phase, and a lack of consistent funding or formalised standards.

\subsubsection{Practitioner Suggestions for Improving Inclusivity}
When reflecting on the challenges of inclusiveness discussed above, participants noted several recurring considerations for how disaster information graphics could better account for vulnerable communities. Rather than focusing only on visual adjustments, interviewees emphasised that inclusiveness is shaped by broader development processes, stakeholder priorities, and structural constraints within emergency communication practice.

\textbf{Types of Graphics: } Participants most frequently identified pictorial action guides and maps as the most important forms of information graphics in disaster communication. Pictorial action guides included step-by-step preparedness visuals and action flow diagrams that illustrate what communities should do before or during an event. Maps were described as central for communicating geographically grounded risk, such as flood-prone area maps, earthquake impact zone maps, and tsunami reach maps. Several participants emphasised the particular importance of map-based communication in real-time warning systems. P5 noted that \textit{``map-based information is central to what we do, but there’s been very little research on making it more inclusive''} (P5), highlighting both its operational value and its ongoing accessibility challenges. Participants also noted that more inclusive scientific visualisations could be valuable in disaster communication. However, P1 emphasised that such graphics need to be presented in an engaging way and may benefit from incorporating a storytelling perspective to support wider community understanding. In addition, evacuation diagrams were also identified as another widely used and high-impact graphic format, and therefore one that would particularly benefit from greater inclusiveness.

\textbf{Co-design and Stakeholders Involvement: }
Participants frequently linked inclusiveness to the extent to which communities are involved in the production of disaster graphics. Several interviewees described the need for sustained engagement and iterative development, particularly when designing for vulnerable groups. P5 noted that meaningful progress depends on the ability to undertake co-design work over time, explaining that \textit{``trying to actually undertake some of that co-design...biggest recommendation at this point...is to spend time in it, rather than money''} (P5). This reflects a broader view that inclusiveness is not achieved through isolated changes, but through ongoing participation and relationship-building.

\textbf{Balance Stakeholder Priorities through Negotiation: }
Participants also highlighted that disaster information graphics often need to satisfy different audiences simultaneously, including emergency authorities, responders, and affected communities. P3 observed that \textit{``different users have different priorities...balancing perspectives requires negotiation and compromise''} (P3). These accounts suggest that inclusive communication is partly shaped by governance questions around whose needs are prioritised and how competing expectations are managed in graphic design decisions.

\textbf{Localisation and Contextual Fit: }
Another recurring theme was that inclusiveness cannot be assumed through universally applied templates. Participants cautioned that graphic guidance may not transfer safely across regions if local built environments, cultural practices, or hazard contexts differ. P2 emphasised the importance of adapting communication materials, referring to the need for \textit{``localised text, instead of just adopting the same steps for all locations''} (P2). This reflects how practitioners associate inclusiveness with context-sensitive design rather than standardisation alone.

\textbf{Address Structural and Resourcing Constraints: }
Finally, participants noted that inclusiveness is also limited by organisational capacity, available guidance, and resourcing. P4 described these barriers as partly structural, pointing to a \textit{``lack of resource...it’s about working out which direction we need to turn to''} (P4). Such perspectives indicate that accessibility gaps persist not only because of design complexity, but also because agencies lack consistent standards, workflows, or institutional support for implementing inclusive practices across preparedness and response phases.

\section{Discussion}
This study examined how information graphics are used in disaster communication, the challenges that shape their effectiveness, and their current inclusivity for vulnerable populations. Overall, the findings confirm that graphics are an essential part in disaster risk communication. Participants described maps, warning visuals, evacuation diagrams, and action-oriented preparedness guides as essential for communicating hazard information quickly and at scale. At the same time, they consistently highlighted that these graphics do not work equally well for all audiences. Inclusiveness remains uneven, and vulnerable communities continue to face barriers in accessing, interpreting, or acting on disaster information graphics.

In this section, we discuss what these findings mean for disaster communication practice, how they extend existing work on disaster risk visualisation and vulnerability, and why they point to the need for new frameworks and technological approaches for more inclusive disaster graphics.

\subsection{Graphics are essential, but accessibility is uneven}
A central insight from this work is that disaster information graphics are widely expected and heavily relied on in time-critical communication. Participants framed maps and warning visuals as central tools for helping communities understand where hazards are, how serious they are, and what actions may be required. Prior studies similarly show that pictures and visual communication can improve attention, comprehension, recall, and adherence, particularly when information must be processed quickly \cite{houts_role_2006,josephson2020handbook,dootson2021managing}. However, our findings also show that the accessibility of these graphics is not universal. Participants raised recurring challenges for vision-impaired users, cognitively impaired audiences, older adults, and people with low literacy or limited familiarity with standard visual formats. In such cases, graphics intended to simplify information can instead become difficult to interpret, particularly under stress. This reinforces findings from disaster warning research showing that visual formats such as maps and warning messages can shape public understanding and responses in uneven ways \cite{liu_is_2017}. Overall, disaster graphics may be central communication instruments, but their benefits are not equally distributed across communities.

\subsection{Inclusiveness is not only a design issue, but a systems issue}
Participants repeatedly emphasised that inclusiveness cannot be treated as a matter of small visual adjustments alone (e.g., colour tweaks or cleaner layouts). Instead, accessibility challenges were linked to deeper structural and operational issues, including limited guidance for complex graphics, uneven engagement with diverse communities, and workflow constraints in emergency settings. This was particularly visible in discussions around maps. Participants described mapped information as one of the most relied-on communication formats, yet also one of the least supported in accessibility frameworks. Several practitioners noted that maps are often treated as a ``too hard'' category, where designers are advised to provide list-based alternatives rather than improving the visual itself.

This aligns with broader literature arguing that disaster risk visualisation must be grounded in user-centred design processes, because interpretation of risk is shaped by psychological, social, and contextual factors \cite{twomlow_user-centred_2022, niyazi2023application, briere2000prevalence}. Disaster graphics, therefore, cannot be evaluated only as static artefacts; they must be understood within the systems through which they are produced, interpreted, and acted upon.

\subsection{Vulnerability is Dynamic}
Another key theme in our findings is that disaster graphics are often treated as reusable templates, even though safe interpretation depends strongly on local and cultural context. Participants cautioned that action guidance may not transfer safely across regions because housing structures, hazard profiles, and everyday practices differ. This reinforces that disaster information graphics are not universal artefacts, but context-sensitive communication tools. Inclusiveness also extends beyond disability-oriented accessibility. Participants raised multilingual barriers, representation concerns, and challenges in reaching culturally and linguistically diverse communities in real time. These findings resonate with recent work on ``digital vulnerability,'' which highlights how uneven digital skills, access, and trust shape communities’ capacity to acquire and act on online risk information \cite{hao_examining_2022,wang2023disaster}. Together, these insights suggest that vulnerability in disaster graphics is multidimensional: it involves disabilities accessibility, cultural and linguistic fit, and socio-technical conditions such as the digital divide.

\subsection{Real-time response constraints create a critical inclusiveness gap}
Participants also made clear that inclusive practices are easier to achieve during preparedness than during real-time response. Agencies may have time in advance to consult communities, test materials, and develop accessible versions. However, during unfolding emergencies, the priority becomes issuing warnings quickly, leaving little capacity to redesign or adapt visuals. This was described as a major barrier for multilingual access and disability-oriented adaptations. Participants also noted that engagement with vulnerable communities is difficult to sustain, and that accessibility work is often underfunded or reactive rather than embedded. These operational constraints point to a central challenge: if inclusive versions of disaster graphics always need to be pre-made, then accessibility becomes difficult to scale across hazards, regions, and diverse needs.

\section{Way Forward and Research Agenda}
The findings of this study highlight that disaster information graphics are essential communication tools, but their inclusiveness remains uneven across vulnerable communities. Participants described inclusivity as a persistent challenge, particularly for map-based and digital-first warning systems, and noted that agencies often lack the capacity to address accessibility needs during real-time emergencies. Building on these insights, we outline key directions for practice and a research agenda to support more inclusive disaster information graphics.

\begin{figure*}
  \includegraphics[width=\textwidth, trim={2cm 8cm 3cm 6cm},clip]{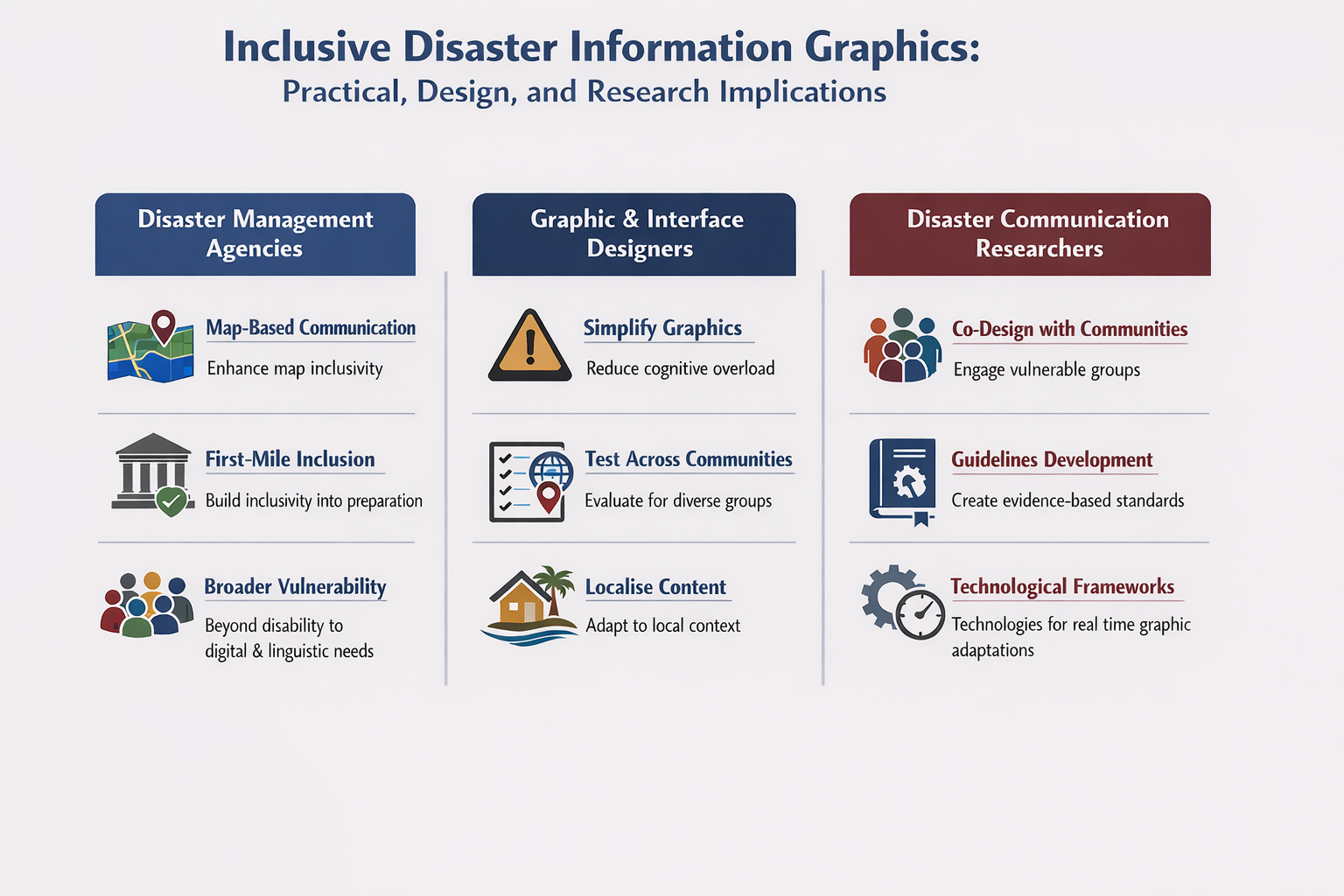}
  \caption{Way Forward for Disaster Communication Graphics}
  \label{issues}
\end{figure*}

\subsection{Implications for disaster management agencies}
For disaster communication agencies, the findings point to several priorities for improving inclusive graphic communication.

\begin{enumerate}
    \item \textbf{Map-based Communication: }Agencies increasingly rely on map-based and digital-first warning systems because they support real-time flexibility. Yet mapped communication remains one of the least accessible formats, and agencies need to explore other options to represent geographic disaster information or encourage initiatives that support developing guidelines to support map inclusivity
    \item \textbf{Inclusivity not as Last-Mile: }Inclusiveness cannot be left to response-phase improvisation which is also known as last-mile solutions. Agencies need to treat inclusive graphic development as part of preparedness infrastructure, embedded into workflows as a first-mile solution rather than handled through ad hoc adaptations.
    \item \textbf{Broadening Vulnerability Definition: }Agencies must recognise that vulnerability is not limited to disability categories alone, but also includes linguistic barriers, digital vulnerability, and unequal access to communication channels. Infographic-based preparedness materials remain useful, but they must be designed with broader inclusiveness in mind \cite{yavar_effective_2012,blake_get_nodate}.
\end{enumerate}

\subsection{Implications for disaster graphic and interface designers}
For designers of disaster information graphics, the findings highlight several priorities.

\begin{enumerate}
    \item \textbf{Complexity of Graphics: }Participants described how layered hazards, dense icons, and complex visual encodings can overwhelm users under stress, even without impairments. Therefore initiatives must be taken to prioritise graphic simplification and cognitive accessibility of these graphics need to be treated as central. This will allow to make graphics more inclusive for several vulnerable communities.
    \item \textbf{More Testing for Inclusivity: }Widely used ``standard'' formats such as charts and map-based dashboards are not universally interpretable. Inclusive design therefore requires interpretability testing across diverse communities, not only aesthetic refinement.
    \item \textbf{Localisation: }Localisation of graphics is critical. Preparedness and action graphics must reflect local housing types, hazard realities, and cultural practices so that the guidance remains safe when applied.
\end{enumerate}

\subsection{Implications for research in disaster communication}
Findings of this study, point to several research gaps and highlight a need for new frameworks and technological solutions that enable disaster information graphics to be adapted quickly and systematically in real time, supporting inclusive communication across hazards, contexts, and diverse community needs. 

\begin{enumerate}
    \item \textbf{Co-design with vulnerable communities: }Participants highlighted that inclusive disaster graphics require deeper engagement with vulnerable communities themselves. While agencies recognise key groups such as vision-impaired, elderly, and multilingual communities, sustained participation and feedback remain difficult to achieve. In some cases, pilots produced only minimal feedback. These findings suggest that inclusiveness cannot be addressed without stronger co-design and requirements gathering processes. Understanding what different groups actually need from disaster information graphics is a necessary foundation for developing practical guidelines.
    \item \textbf{Guidelines for inclusive disaster graphics: } Another recurring issue was the lack of practical, evidence-based guidance on how disaster graphics should be adapted for different vulnerable communities. Existing standards remain limited for complex formats such as hazard maps. Participants of the study specifically mentioend that more work in the space to develop more guidance on graphics will be highly appreciated. This indicates the need for stronger guidelines that specify what aspects of disaster graphics should change, and how, for different vulnerable groups.
    \item \textbf{Technological frameworks for real-time adaptation: }
    Most importantly, participants repeatedly noted that agencies are often not able to make graphics accessible during an unfolding disaster. This creates a critical challenge. If accessible versions of graphics always need to be pre-made, then inclusiveness becomes difficult to scale across hazards, regions, and diverse community needs. This highlights the need for frameworks and technological approaches that can support more systematic and real-time adaptation of disaster information graphics.
\end{enumerate}

Overall, this study reinforces that disaster graphics are essential communication infrastructure, but also that inclusiveness remains uneven and constrained by both design limitations and emergency operational realities. Addressing these gaps requires advances in co-design practice, clearer accessibility guidance, and technological frameworks that can support inclusive graphics at the pace disasters demand.

\section{Limitations}
This study provides practitioner perspectives on the use and inclusiveness of disaster information graphics. While the findings offer valuable insights into current practices and challenges, several limitations should be acknowledged.

\textbf{Limited sample size and participant scope: }
The interview study involved a small number of disaster communication and emergency management practitioners. Although participants brought substantial expertise and offered rich accounts of real-world graphic use, the findings may not represent the full diversity of disaster agencies, jurisdictions, or communication settings. Future work with a larger and broader sample, including practitioners from different regions and organisational contexts, would strengthen generalisability.

\textbf{Practitioner-centred perspective: }
This study focused on expert and agency perspectives rather than directly capturing the experiences of vulnerable community members themselves. While practitioners provided reflections on accessibility gaps and engagement challenges, the needs and interpretations of vulnerable populations may differ. Further studies involving disabled communities, culturally and linguistically diverse groups, and other vulnerable audiences are necessary to validate and extend these findings from an end-user perspective.

\textbf{Contextual and geographic specificity: }
Most participants worked within Australian disaster communication systems, where digital-first warning platforms and agency structures shape how information graphics are produced and disseminated. As a result, some findings may reflect local operational arrangements, hazard types, or policy environments. Disaster communication practices may differ across countries, particularly in regions with different levels of infrastructure, literacy, or emergency governance.

\textbf{Data interpretation and qualitative coding: }
As with qualitative research, analysis involved interpretation of participant responses and thematic coding. Although quotes were used to ground themes in the data, different researchers may emphasise different aspects of the transcripts. We addressed this threat by one researcher conducting the coding and verifying these codes. However, to reduce subjectivity, future work could incorporate triangulation with complementary data sources such as graphic artefact analysis or observational studies.


\textbf{Rapidly evolving tools and practices: }
Finally, disaster communication technologies and accessibility standards continue to evolve. The practices described by participants reflect current organisational constraints, but future advancements in adaptive communication tools, multilingual platforms, or accessibility guidance may shift how inclusiveness is addressed over time.

Despite these limitations, the study provides important early evidence of how disaster information graphics are currently designed, interpreted, and constrained in practice, and highlights key gaps that motivate further work on inclusive graphic frameworks.

\section{Conclusion}

Information graphics play a central role in disaster communication, supporting preparedness, warning dissemination, and real-time response. Through interviews with disaster communication practitioners, this study examined how such graphics are currently used, what challenges shape their effectiveness, and how inclusive they are for vulnerable populations. Our findings show that graphics such as maps and pictorial action guides are widely relied on as essential communication tools. However, participants highlighted that these graphics do not work equally well for all audiences. Vision-impaired users, cognitively impaired communities, older adults, culturally and linguistically diverse groups, and people with low literacy may face significant barriers when graphics are visually dense, cognitively demanding, culturally mismatched, or dependent on digital access. Importantly, participants emphasised that inclusiveness is not only a design challenge, but also an operational and systemic one. Agencies described limited capacity to adapt graphics during unfolding emergencies, uneven community engagement, and a lack of clear guidance for making complex formats such as hazard maps accessible in practice. These constraints mean that accessibility efforts often remain fragmented, reactive, or restricted to what can be prepared in advance. This study highlights three key directions for future work: (1) stronger co-design and requirements gathering with vulnerable communities to better understand inclusiveness needs, (2) evidence-based guidelines that clarify what should change in disaster information graphics for different groups, and (3) technological frameworks that can support real-time accessibility of disaster graphics during emergency response. Overall, this work positions inclusive disaster information graphics as essential to communication infrastructure and motivates the need for more systematic and community-driven approaches to ensure that disaster warnings leave no one behind.

\section*{Acknowledgments}
Grundy and Xiao are supported by ARC Laureate Fellowship FL190100035




\bibliographystyle{elsarticle-num-names} 
\bibliography{main}

\end{document}